\pdfoutput=1
\documentclass[twocolumn]{aastex62}
\usepackage{amsmath}
\received{\today}
\submitjournal{ApJL}

\shorttitle{Testing a correlation between UHE cosmic rays and starburst galaxies}
\shortauthors{Telescope Array collaboration}

\begin{document}

\title{Testing a reported correlation between arrival directions of ultrahigh-energy cosmic rays and a flux pattern from nearby starburst galaxies
using Telescope Array data}

\correspondingauthor{Armando di Matteo, Toshihiro Fujii, Kazumasa Kawata}
\email{armando.di.matteo@ulb.ac.be, fujii@icrr.u-tokyo.ac.jp, kawata@icrr.u-tokyo.ac.jp}

\collaboration{Telescope Array collaboration}
\author[0000-0001-6141-4205]{R.U. Abbasi}
\affiliation{High Energy Astrophysics Institute and Department of Physics and Astronomy, University of Utah, Salt Lake City, Utah, USA}

\author{M. Abe}
\affiliation{The Graduate School of Science and Engineering, Saitama University, Saitama, Saitama, Japan}

\author[0000-0001-5206-4223]{T. Abu-Zayyad}
\affiliation{High Energy Astrophysics Institute and Department of Physics and Astronomy, University of Utah, Salt Lake City, Utah, USA}

\author{M. Allen}
\affiliation{High Energy Astrophysics Institute and Department of Physics and Astronomy, University of Utah, Salt Lake City, Utah, USA}

\author{R. Azuma}
\affiliation{Graduate School of Science and Engineering, Tokyo Institute of Technology, Meguro, Tokyo, Japan}

\author{E. Barcikowski}
\affiliation{High Energy Astrophysics Institute and Department of Physics and Astronomy, University of Utah, Salt Lake City, Utah, USA}

\author{J.W. Belz}
\affiliation{High Energy Astrophysics Institute and Department of Physics and Astronomy, University of Utah, Salt Lake City, Utah, USA}

\author{D.R. Bergman}
\affiliation{High Energy Astrophysics Institute and Department of Physics and Astronomy, University of Utah, Salt Lake City, Utah, USA}

\author{S.A. Blake}
\affiliation{High Energy Astrophysics Institute and Department of Physics and Astronomy, University of Utah, Salt Lake City, Utah, USA}

\author{R. Cady}
\affiliation{High Energy Astrophysics Institute and Department of Physics and Astronomy, University of Utah, Salt Lake City, Utah, USA}

% ...

\begin{abstract}

The Pierre Auger Collaboration (Auger) recently reported a
correlation between the arrival directions of cosmic rays 
with energies above 39~EeV and the flux pattern of 23
nearby starburst galaxies (SBGs).
In this Letter, we tested the same hypothesis using
cosmic rays detected by the Telescope Array experiment (TA) in the 9-year period
from May~2008 to May~2017. 
Unlike the Auger analysis, we did not optimize the parameter values 
but 
kept them fixed to the best-fit values found by Auger, namely
9.7\% for the anisotropic fraction of cosmic rays assumed to originate from the
SBGs in the list and $12.9^\circ$ for the angular scale of the
correlations.  The energy threshold we adopted is 43~EeV, 
corresponding to 39~EeV in Auger when taking into account
the energy-scale difference between two experiments.
We find that the TA data is compatible with isotropy to within $1.1\sigma$ 
and with the Auger result to within $1.4\sigma$, meaning that it is not capable to discriminate between these two hypotheses.

\end{abstract}

\keywords{astroparticle physics --- cosmic rays --- galaxies: starburst --- methods: data analysis}

\section{Introduction} \label{sec:intro}

The origins of ultrahigh-energy cosmic rays (UHECRs) are still unknown.
Anisotropies in the angular distribution of their arrival directions are
rather small, requiring the detection of a large number of events to observe them. 
Furthermore, deflections of UHECRs by Galactic and intergalactic
magnetic fields complicate the interpretation of anisotropies in terms of possible
sources; this effect is reduced 
for the highest-energy cosmic rays, but the available statistics are significantly limited 
due to the steeply falling spectrum of UHECRs.

The two largest UHECR observatories in operation are the Telescope Array
\citep[hereinafter TA,][]{AbuZayyad:2012kk}, located in Utah, USA,
with approximately 700~km$^2$ effective area, and the Pierre Auger Observatory
\citep[hereinafter Auger,][]{ThePierreAuger:2015rma}, located in Argentina
with 3000~km$^2$ effective area. Their exposures peak in the Northern and
Southern hemispheres, respectively.

Auger recently reported \citep{Aab:2018chp} a correlation
between UHECR events with reconstructed energies above 39~EeV and a flux pattern 
of nearby starburst galaxies (SBGs). A model where $90.3\%$ of the flux is
isotropic and $9.7\%$ originates from SBGs (with UHECR luminosities assumed
proportional to their radio luminosities) and undergoes Gaussian random
deflections with standard deviation $12.9^\circ$ in each transverse dimension
is favored over the purely isotropic model with a post-trial significance of
4.0$\sigma$,
and over a model based on the overall galaxy distribution beyond 1~Mpc with a
$3.0\sigma$ significance. In the Auger analysis it was found that
different selections of candidate sources yield very similar results, as in
any case over $90\%$ of the anisotropic part of the flux weighed by the Auger
directional exposure originates from four bright objects --- NGC~4945, NGC~253,
M83, and NGC~1068.

In this Letter, we follow up on this finding by testing UHECRs detected by TA 
in the Northern hemisphere against the same flux model and the best-fit values reported by Auger, 
and discuss possible interpretations of our result.

\section{Analysis}
\label{sec:analysis}
\subsection{Cosmic-ray dataset}
\label{sec:cosmic-ray-dataset}

The TA is located at $39.3^\circ$~N, $112.9^\circ$~W, in Millard County, Utah, USA, about 200~km
south-west of Salt Lake City, about 1400~m above sea level \citep{AbuZayyad:2012kk}.
The TA surface detector (SD) array consists of 507 plastic scintillation detectors 
on a square grid with 1.2~km spacing, covering an area of 700~km$^2$, 
and is surrounded by three fluorescence detector (FD)
stations \citep{Tokuno:2012mi} with telescopes overlooking the SD array. It has been collecting data
since May~2008. The SD has $\approx 100\%$ duty cycle, against $\approx 10\%$
for the FD, so with a similar collection area the SD has about ten times
the statistics. The events detected in coincidence by both detectors
are used to calibrate energy scale of the SD:
SD reconstructed energies (determined by comparison to Monte Carlo simulations)
are rescaled by a factor of 1/1.27 to match the FD energy scale (determined calorimetrically) \citep{AbuZayyad:2012ru,spectrum2017}.
The systematic uncertainty on the TA energy scale is 21\% \citep{TheTelescopeArray:2015mgw} and its energy and angular resolutions are
$15$--$20\%$ and $1.0$--$1.5^\circ$, respectively, depending on the event
geometry and energy \citep{Abbasi:2014lda}. 

In this work we use data collected by the TA SD array in a
9-year period from May~2008 to May~2017 with reconstructed
energies above 43~EeV, zenith angles less than $55^\circ$,
and declinations~$\delta > -10^\circ$ using the same
quality cuts as in~\citet{Abbasi:2014lda}.  This dataset comprises 284 events.
We neglect the finite angular and energy resolution of TA events, and consider
the detector fully efficient, i.e.~with a flat response for all showers
with energies and zenith angles in the considered range, so that its directional exposure $\omega_\text{TA}$ equals the geometrical one
for $\delta > -10^\circ$,
which varies with declination but not with right ascension \citep{Sommers:2000us}:
\begin{align}
	\omega_\text{TA}(\delta)	&\propto \cos\phi_\text{TA} \cos\delta \sin\alpha_\text{m} + \alpha_\text{m} \sin\phi_\text{TA} \sin\delta, 
	\label{eq:expo}\\
	\alpha_\text{m}	&= \begin{cases}
		\pi,	& \xi < -1; \\
		\arccos \xi, & -1 \le \xi \le 1; \\
		0,	& \xi > 1;
	\end{cases} \nonumber\\
	\xi	&= \frac{\cos\theta_\text{m} - \sin\phi_\text{TA} \sin\delta}{\cos\phi_\text{TA} \cos\delta},\nonumber
\end{align}
where $\phi_\text{TA} = +39.3^\circ$ is the detector latitude and $\theta_\text{m} = 55^\circ$ is the maximum zenith angle accepted.

The energy threshold of $E_{\min} = 43$~EeV used in this analysis corresponds
to the Auger energy threshold of $39$~EeV at which the most significant
correlation with SBG was found. Here we took into
account the $10.4\%$ difference between the energy scales of the
two experiments as estimated by a comparison of energy spectra around 5 EeV \citep{AbuZayyad:2018aua,Verzi:2017hro}.

\subsection{Source catalog}\label{sec:sources}
Following the Auger analysis \citep{Aab:2018chp}, we select the candidate sources from a sample of
63 SBGs outside the Local Group compiled by the \textit{Fermi}-LAT
collaboration \citep{Ackermann:2012vca} for the gamma-ray emission
search.\footnote{Only four of those objects were actually successfully
  detected in gamma rays in that work: NGC~253, M82, NGG~4945 and NGC~1068.}
Imposing the cut of flux greater than 0.3~Jy at 1.4~GHz leaves 23 objects in the catalog of
candidate sources. Their UHECR fluxes were assumed to be proportional to
their radio fluxes at 1.4~GHz. These objects are listed in
Table~\ref{table:SBG}.

In the Auger analysis, the effect of energy losses by UHECRs during their propagation was found to be negligible in the SBG model, as most of the anisotropic flux originates from sources within a
few~Mpc; in this work, we neglected the losses for simplicity.
\begin{table}
\caption{Selected source candidates from the
SBG catalog used in this analysis (the same as in
  \citealt{Aab:2018chp}). The last column shows the relative source contribution
  weighted with the TA directional exposure $\omega_\text{TA}$. }
\label{table:SBG}
\begin{tabular}{crrrrr}
\hline \hline
name & \multicolumn{2}{c}{Gal.~$(l,b)$} & distance & \multicolumn{1}{c}{flux $\phi$} & $\phi\omega_\text{TA}$ \\
\hline
NGC~253 & $97.4^\circ$ & $-88.0^\circ$ & $2.7$~Mpc & $13.6\%$ & $1.6\%$ \\
M82 & $141.4^\circ$ & $40.6^\circ$ & $3.6$~Mpc & $18.6\%$ & $35.7\%$ \\
NGC~4945 & $305.3^\circ$ & $13.3^\circ$ & $4.0$~Mpc & $16.0\%$ & $0.0\%$ \\
M83 & $314.6^\circ$ & $32.0^\circ$ & $4.0$~Mpc & $6.3\%$ & $0.4\%$ \\
IC~342 & $138.2^\circ$ & $10.6^\circ$ & $4.0$~Mpc & $5.5\%$ & $10.5\%$ \\
NGC~6946 & $95.7^\circ$ & $11.7^\circ$ & $5.9$~Mpc & $3.4\%$ & $6.2\%$ \\
NGC~2903 & $208.7^\circ$ & $44.5^\circ$ & $6.6$~Mpc & $1.1\%$ & $1.4\%$ \\
NGC~5055 & $106.0^\circ$ & $74.3^\circ$ & $7.8$~Mpc & $0.9\%$ & $1.5\%$ \\
NGC~3628 & $240.9^\circ$ & $64.8^\circ$ & $8.1$~Mpc & $1.3\%$ & $1.5\%$ \\
NGC~3627 & $242.0^\circ$ & $64.4^\circ$ & $8.1$~Mpc & $1.1\%$ & $1.2\%$ \\
NGC~4631 & $142.8^\circ$ & $84.2^\circ$ & $8.7$~Mpc & $2.9\%$ & $4.4\%$ \\
M51 & $104.9^\circ$ & $68.6^\circ$ & $10.3$~Mpc & $3.6\%$ & $6.2\%$ \\
NGC~891 & $140.4^\circ$ & $-17.4^\circ$ & $11.0$~Mpc & $1.7\%$ & $2.8\%$ \\
NGC~3556 & $148.3^\circ$ & $56.3^\circ$ & $11.4$~Mpc & $0.7\%$ & $1.3\%$ \\
NGC~660 & $141.6^\circ$ & $-47.4^\circ$ & $15.0$~Mpc & $0.9\%$ & $1.0\%$ \\
NGC~2146 & $135.7^\circ$ & $24.9^\circ$ & $16.3$~Mpc & $2.6\%$ & $5.2\%$ \\
NGC~3079 & $157.8^\circ$ & $48.4^\circ$ & $17.4$~Mpc & $2.1\%$ & $3.8\%$ \\
NGC~1068 & $172.1^\circ$ & $-51.9^\circ$ & $17.9$~Mpc & $12.1\%$ & $9.1\%$ \\
NGC~1365 & $238.0^\circ$ & $-54.6^\circ$ & $22.3$~Mpc & $1.3\%$ & $0.0\%$ \\
Arp~299 & $141.9^\circ$ & $55.4^\circ$ & $46.0$~Mpc & $1.6\%$ & $2.9\%$ \\
Arp~220 & $36.6^\circ$ & $53.0^\circ$ & $80.0$~Mpc & $0.8\%$ & $1.1\%$ \\
NGC~6240 & $20.7^\circ$ & $27.3^\circ$ & $105.0$~Mpc & $1.0\%$ & $0.8\%$ \\
Mkn~231 & $121.6^\circ$ & $60.2^\circ$ & $183.0$~Mpc & $0.8\%$ & $1.4\%$ \\
\hline
\end{tabular}
\end{table}

\newcommand{\n}{\hat{\mathbf{n}}}
\newcommand{\skyint}[1]{\int_{4\pi} #1 \,\mathrm{d}\Omega}
\newcommand{\norm}[1]{\frac{#1}{\skyint{#1}}}
\newcommand{\TS}{\mathrm{TS}}
\newcommand{\Piso}{\Phi_\text{iso}}
\newcommand{\fSBG}{f_\text{SBG}}
\subsection{Test statistic and flux model}
Let $\n$ be the unit vector representing a direction in the sky, pointing away from the observer.
Given two flux models $\Phi_1(\n), \Phi_2(\n)$ describing a null hypothesis and an alternative hypothesis, respectively,
and the directional exposure~$\omega(\n)$ of an experiment,
the test statistic (hereinafter $\TS$) is defined as twice the log-likelihood ratio
\begin{align}
	\TS		&= 2 \ln \left(L(\Phi_2) / L(\Phi_1)\right), \label{eq:TS}\\
	\text{where}~L(\Phi_j)	&= \prod_i \frac{\Phi_j(\n_i)\omega(\n_i)}{\skyint{\Phi_j(\n)\omega(\n)}}, \nonumber
\end{align}
and $\n_i$ being the reconstructed arrival direction of the $i$-th observed event.
A positive (negative) $\TS$ indicates that the dataset is more (less) likely if the
real flux is described by~$\Phi_2(\n)$ than by~$\Phi_1(\n)$.

\begin{figure*}
\plottwo{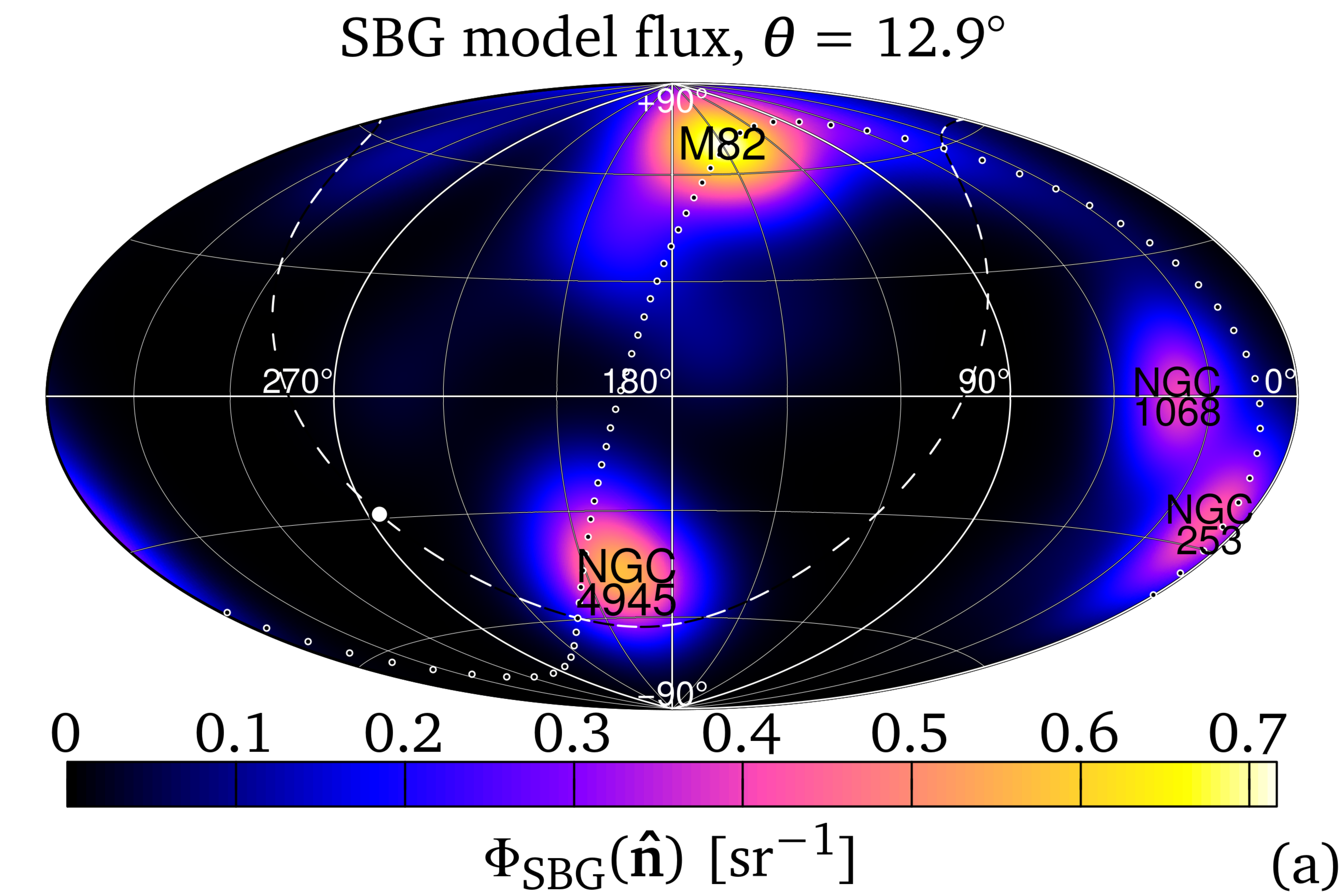}{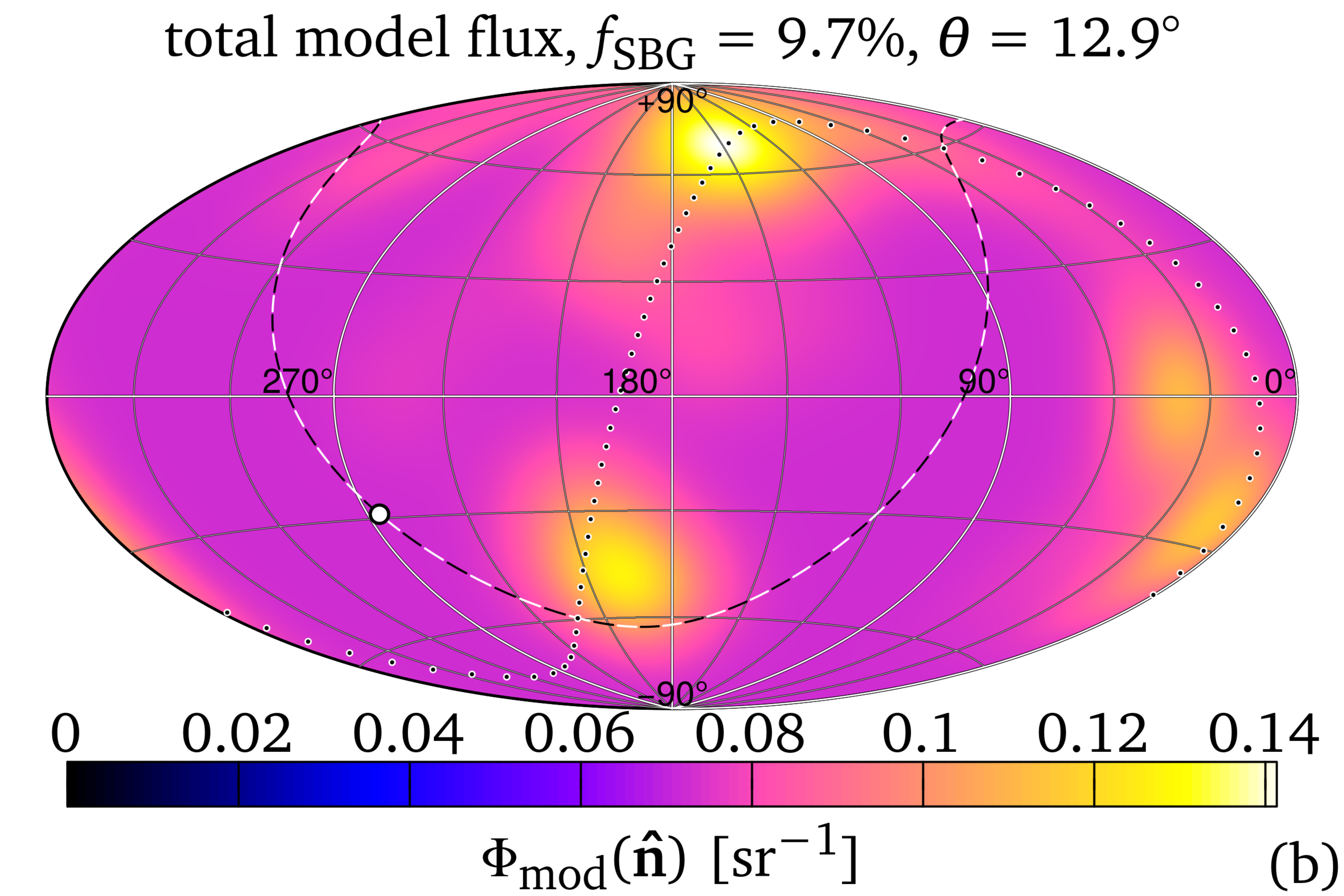}
\plottwo{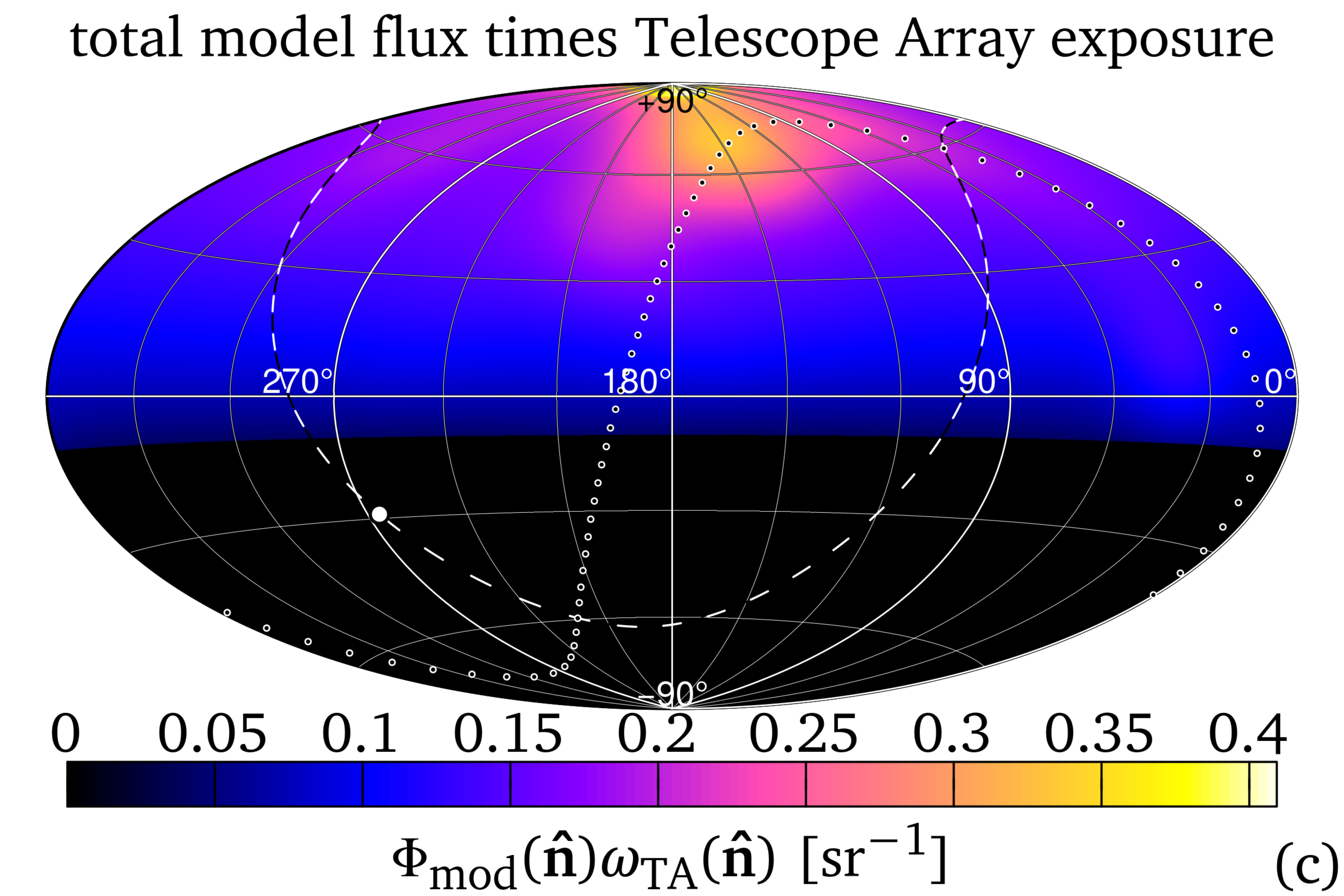}{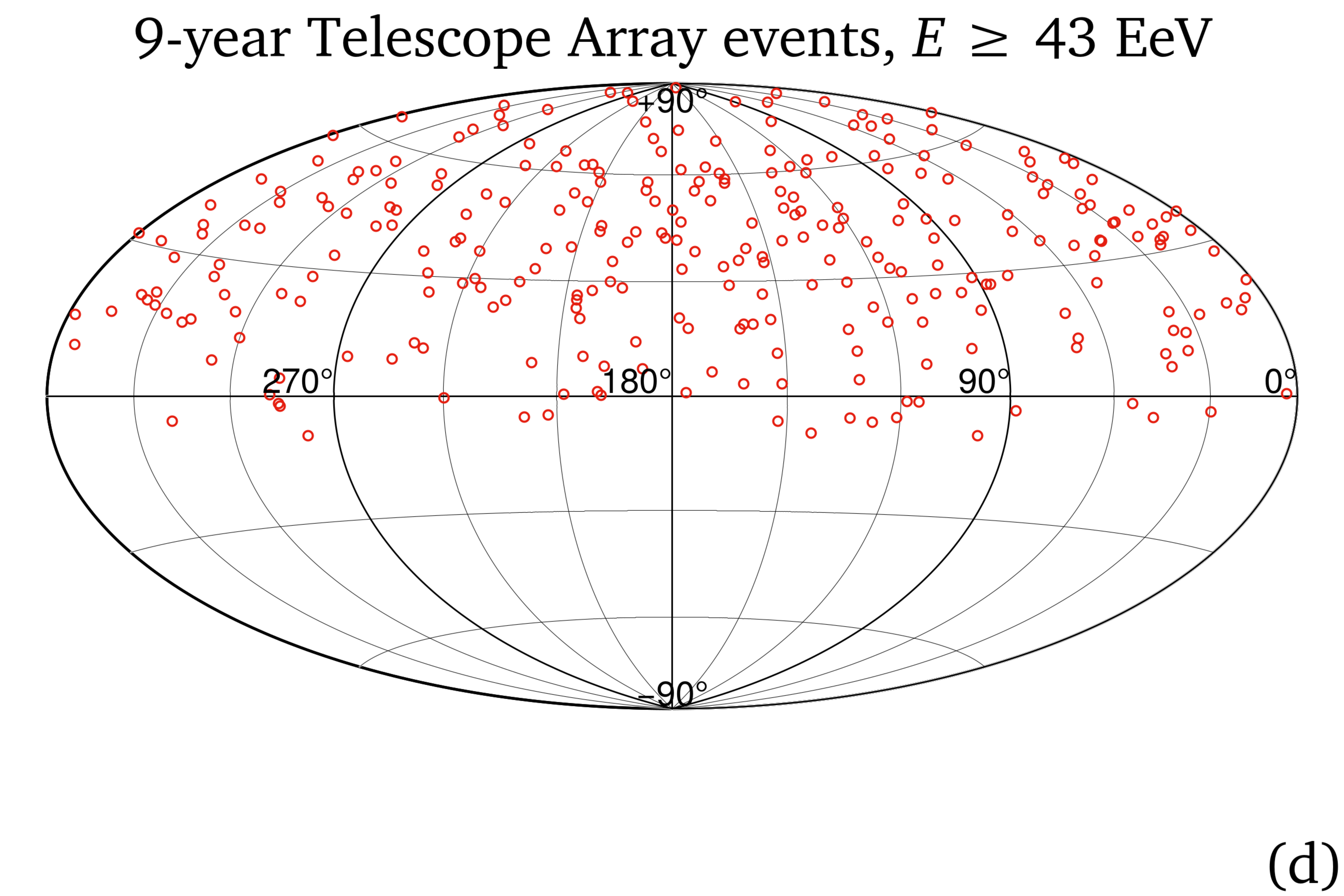}
\caption{Maps of: (a) the anisotropic part of the model flux (equation~\ref{eq:Phi_SBG}); (b) the total model flux (equation~\ref{eq:Phi_mod});
(c) the total model flux multiplied by the TA exposure; and (d) the TA events above 43~EeV.
The dashed and dotted lines represent the Galactic and supergalactic planes respectively and the white disk shows the Galactic center.}
\label{fig:maps}
\end{figure*}

In this analysis, the null hypothesis is an isotropic flux, $\Phi_1(\n) = \Piso = 1/4\pi$,
whereas the alternative hypothesis is $\Phi_2(\n)$ = \begin{equation}
	\Phi_\text{mod}(\n) = \fSBG \Phi_{\text{SBG}}(\n) + (1-\fSBG) \Piso, \label{eq:Phi_mod}
\end{equation} where $\fSBG = 9.7\%$~is the fraction of the flux assumed to originate from the SBGs in the catalog
(the rest being assumed to be isotropic),
and \begin{equation}
	\Phi_{\text{SBG}}(\n) = \norm{\sum_k \phi_k \exp\left(\n_k\cdot\n/\theta^2\right)} \label{eq:Phi_SBG}
\end{equation}
is a weighed sum of von~Mises--Fisher distributions (the spherical analog of the Gaussian distribution),
where $\phi_k$ and $\n_k$ are the flux and position of the $k$-th source from Table~\ref{table:SBG}
and $\theta=12.9^\circ$~is the RMS deviation in each transverse dimension,
the total RMS deviation being $\sqrt{2}\theta$. The exposure is assumed to be geometrical, $\omega(\n) = \omega_\text{TA}(\n)$ from
eq.~\eqref{eq:expo}.
In the present work we do not optimize the parameter values 
but keep them fixed to the Auger best-fit values,
in order not to include any freedom in the model which would require a
statistical penalty.
The resulting model flux is shown in Figure~\ref{fig:maps}, along with the events in the TA dataset.

\section{Results}
Substituting the coordinates of the TA events $\{\n_i\}$ into equation~(\ref{eq:TS}), the test statistic we obtained was~$\TS = -1.00$.
In order to assess the significance of this result, we computed TS for
$10^6$ Monte~Carlo (MC) datasets generated assuming an isotropic flux, and found~$\TS \ge -1.00$
in~$p=14.3\%$ of the $10^6$ cases, corresponding to a $1.1\sigma$~significance.%
\footnote{Note that unlike 
in the Auger analysis, Wilks' theorem is not applicable here because we did not scan a parameter space which the null hypothesis is a subspace of.}

\begin{figure}
  \plotone{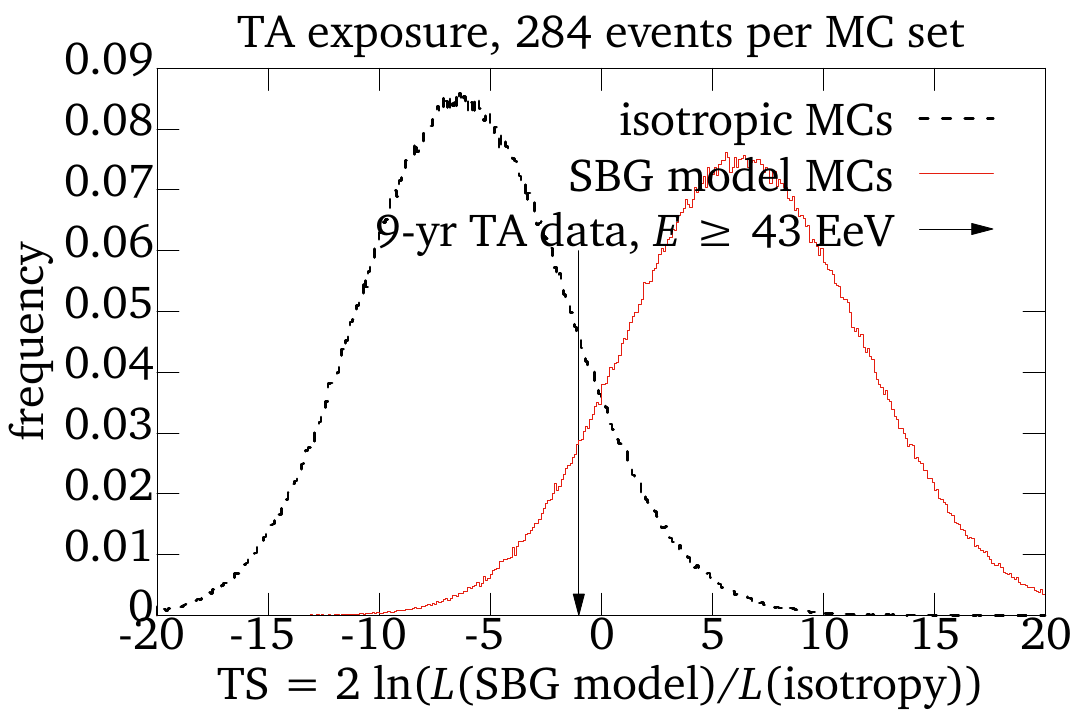}
  \caption{Distribution of test statistics in MC sets generated according to
  the two flux hypotheses we considered.}
  \label{fig:MC_TS}
\end{figure}

We also computed test statistics for $10^6$ MC sets generated under an assumption of the Auger best-fit SBG flux model
to know the range of $\TS$ values that could be expected in that case.
The results are shown in Figure~\ref{fig:MC_TS}.
We found that $92.5\%$ of realizations in the latter case have a higher $\TS$ value
than the TA data (corresponding to a $-1.4\sigma$~significance).
We also verified that, as should be by design, the ratio between the two $\TS$~distributions is $\exp(\TS/2)$.
A negative $\TS$ means that the angular distribution in a dataset resembles isotropy more than the SBG model,
and a positive $\TS$ means the reverse,
so most isotropic realizations have $\TS < 0$ and most SBG-like realizations have $\TS > 0$.
$\TS \approx 0$ would mean that the angular distribution in a dataset is about equally different
from the two models considered.

\section{Discussion}
A limitation in this analysis is the exclusion of Local Group objects
(SMC, LMC, M33 and M31), which were listed in \citet{Ackermann:2012vca}, but in
a separate table.  
These objects are not particularly intrinsically luminous (several times less
than the dimmest objects in Table~\ref{table:SBG}),
but due to their proximity ($D=0.06$, $0.05$, $0.85$ and $0.78$~Mpc respectively)
they appear very bright. If the assumed proportionality between the UHECR luminosity,
the star-formation rate and the radio luminosity also applied to them, then the LMC and
SMC would outshine all other objects combined in the Auger sky, and M33 and
M31 would be the second and third brightest objects in the TA
sky; but no excess of events is apparent in the vicinity of either pair of objects
in our data or in \citet{Aab:2018chp}.
A discussion about possible theoretical astrophysical motivations for not including these objects
in the sample is outside the scope of this work.

\citet{Aab:2018chp} also tested their data for correlations
with gamma-ray loud AGNs from the 2FHL catalog
\citep{Ackermann:2015uya}. The best fit ($E_{\min}=60$~EeV,
$f_{\gamma\text{AGN}}=6.7\%$, $\theta=6.9^\circ$) is favored over isotropy
at the $2.7\sigma$ level.  
Unlike with SBGs, UHECR energy losses in propagation are not negligible in this case
because the unattenuated flux is not dominated by nearby objects.
Testing TA data for correlations with this catalog would not be very useful,
because the attenuated flux at Earth is dominated by Cen~A, way outside the TA field of view (at
$\delta = -43^\circ$), leaving the flux in the northern hemisphere very
nearly isotropic, and therefore requiring a very large number of events
for an experiment in the northern hemisphere to detect the correlation;
also, the Auger best-fit energy threshold found with this catalog
($E_{\min}=60$~EeV) was higher than with the SBGs, further reducing the available statistics.

\section{Conclusions}
\label{sec:concl}
This Letter presents the result of a search for a correlation between
arrival directions of UHECRs observed by TA and the flux pattern of SBGs.
The SBG sample, anisotropic fraction and angular scale were fixed to be the best-fit values as in the Auger study.
The energy threshold of $43$~EeV was determined by taking into account of the energy scale difference between two experiments
\citep{AbuZayyad:2018aua},
corresponding to $39$~EeV at which the most significant correlation was reported in Auger. 
The result of this test was inconclusive, being compatible both
with isotropy to within $1.1\sigma$ and with the Auger result to within $1.4\sigma$.
This means that the current TA data is not capable to discriminate between these two hypotheses.
The ongoing expansion of TA \citep{Kido:2018dhc} will increase its effective
area by a factor of 4, allowing us to reduce the statistical uncertainties and possibly
to discriminate between different hypothesis about the UHECR origin.
\section*{Acknowledgments}
We thank Pierre Auger collaboration members Jonathan Biteau and Olivier Deligny
for useful discussions about their analysis.


The Telescope Array experiment is supported by the Japan Society for
the Promotion of Science(JSPS) through 
Grants-in-Aid
for Priority Area
431,
for Specially Promoted Research 
JP21000002, 
for Scientific  Research (S) 
JP19104006, 
for Specially Promoted Research 
JP15H05693, 
for Scientific  Research (S)
JP15H05741 and
for Young Scientists (A)
JPH26707011; 
by the joint research program of the Institute for Cosmic Ray Research (ICRR), The University of Tokyo; 
by the U.S. National Science
Foundation awards PHY-0601915,
PHY-1404495, PHY-1404502, and PHY-1607727; 
by the National Research Foundation of Korea
(2016R1A2B4014967, 2016R1A5A1013277, 2017K1A4A3015188, 2017R1A2A1A05071429) ;
by the Russian Academy of
Sciences, RFBR grant 16-02-00962a (INR), IISN project No. 4.4502.13,
and Belgian Science Policy under IUAP VII/37 (ULB). The foundations of
Dr. Ezekiel R. and Edna Wattis Dumke, Willard L. Eccles, and George
S. and Dolores Dor\'e Eccles all helped with generous donations. The
State of Utah supported the project through its Economic Development
Board, and the University of Utah through the Office of the Vice
President for Research. The experimental site became available through
the cooperation of the Utah School and Institutional Trust Lands
Administration (SITLA), U.S. Bureau of Land Management (BLM), and the
U.S. Air Force. We appreciate the assistance of the State of Utah and
Fillmore offices of the BLM in crafting the Plan of Development for
the site.  Patrick Shea assisted the collaboration with valuable advice 
on a variety of topics. The people and the officials of Millard County, 
Utah have been a source of
steadfast and warm support for our work which we greatly appreciate. 
We are indebted to the Millard County Road Department for their efforts 
to maintain and clear the roads which get us to our sites. 
We gratefully acknowledge the contribution from the technical staffs of
our home institutions. An allocation of computer time from the Center
for High Performance Computing at the University of Utah is gratefully
acknowledged.

\end{document}